\documentstyle[epsfig,preprint,aps]{revtex}
\tightenlines 
\input NATOpaper.def
 
\gdef\journal#1, #2, #3, 1#4#5#6{		
    {\sl #1~}{\bf #2}, #3 (1#4#5#6)}		

\def\jcp{\journal J. Chem. Phys., }

\def\prl{\journal Phys. Rev. Lett., }

\def\onlinecite{\cite}
\def\openone{\leavevmode\hbox{\small1\kern-3.8pt\normalsize1}}%
\newcommand{\keyword}[1]{\index{#1}#1}

\begin{document}
\draft
\title{Quantum Monte Carlo Methods in Statistical Mechanics}
\author{Vilen Melik-Alaverdian and M.P. Nightingale}
\address{Department of Physics, University of Rhode Island, Kingston, Rhode Island 02881, USA} 

\date{\today}
\maketitle

\begin{abstract}
This paper deals with the optimization of trial states for the
computation of dominant eigenvalues of operators and very large matrices.
In addition to preliminary results for the energy spectrum of van der
Waals clusters, we review results of the application of this method
to the computation of relaxation times of independent relaxation modes
at the Ising critical point in two dimensions.
\end{abstract}

\section{Introduction}
The computation of eigenvalues and eigenstates of operators and large
matrices is a ubiquitous problem.  In this paper we review recent
applications of the Quantum \MC\ methods that we have developed for
this purpose.  The reader is referred to other papers for introductory
or more technical discussions of earlier
work.\cite{NigUmrAdv99,NATO-ASI98,NB96.prl,NB98.prl}

\section{Mathematical Preliminaries}
For an operator $G$, the power method can be used to compute  the
dominant eigenstate and eigenvalue, $\ket{\psi_0}$ and $\lambda_0$.

This well-know procedure can be summarized as follows:
\begin{enumerate}
\item
Choose a generic initial state $\ket {u^{(0)}}$
of the appropriate symmetry.
\item
Iterate:
\beq
\ket {u^{(t+1)}} = {1\over c_{t+1}}G \ket {u^{(t)}},
\eeq
where $c_t$ puts $\ket {u^{(t)}}$ in standard form.
\end{enumerate}
For projection time $t\to \infty$ the following is almost always true:
\begin{enumerate}
\item
Eigenstate:
\beq
\ket {u^{(t+1)}} \to  \ket{\psi_0}
\eeq
\item
Eigenvalue:
\beq
c_t \to \lambda_0
\eeq
\end{enumerate}
To see this, expand the initial state in normalized eigenstates
\beq
\ket {u^{(0)}}=\sum_k w^{(0)}_k \ket{\psi_k}
\eeq
with spectral weights $w^{(0)}_k$. Then
$\ket {u^{(t)}}$ has spectral weights
\beq
w^{(t)}_k \sim \left({\lambda_k\over\lambda_0}\right)^t.
\eeq

This method can be implemented by means of a \MC\ method and, unlike
variational \MC , it has the advantage of producing unbiased results for
large projection times $t$. The disadvantage is, however, that at the
same time the statistical noise increases exponentially, unless $G$
is a Markov (stochastic) matrix, or can be explicitly transformed
to one.  The statistical errors grow with the extent to which $G$
fails to conserve probility, and to alleviate this problem,
approximate dominant eigenstates can be used.

In the case of Markov matrices, computation of the dominant eigenvalue
is of no interest, since it is equals unity, but sampling the
corresponding eigenstate has numerous applications.

\subsection{Subspace iteration}\index{subsection iteration}
\label{sec.subspace}

Given a set of basis states, one can construct trial states as linear
combinations to obtain approximate excited or, more generally,
sub-dominant states and the corresponding eigenvalues.  These are
computed by solving a linear variational problem.  In a \MC\ context,
the Metropolis method can be used to evaluate the required matrix
elements.  Subsequently, a variation of the power method can again be
used to remove systematically the variational
bias.\cite{CeperleyBernu88,BerCepLes90,BrGlLes95} Again, the price to be paid
for reduction of the variational bias is increased statistical noise,
a problem that can be mitigated by the use of {\it optimized}
trial states.

The linear variational problem to be solved for the computation of
excited states is the following one.  Given $n$ basis functions
$\ket{u_i}$, find the $n\times n$ matrix of coefficients $D_i^{(j)}$
such that
\beq
\ket{\tilde \psi_j} \,=\, \sum_{i\,=\,1}^n D_i^{(j)}\/\ket{ u_i}
\label{eq.varexc}
\eeq
are the ``best'' variational approximations for the $n$ lowest
eigenstates $\ket{\psi_i}$ of some Hamiltonian $\Ham$. In this problem
we shall, at least initially, use the language of quantum mechanical
systems, where one has to distinguish the Hamiltonian from the
imaginary-time evolution operator $G=\exp(-\tau \Ham)$.  In the
statistical mechanical application discussed below, we shall encounter
only the equivalent of the latter, which is the Markov matrix
governing the stochastic dynamics.  In the expressions to be derived
below, the substitution $\Ham G^p \to G^{p+1}$ will produce the
expressions required for the statistical mechanical application, at
least if we assume that the non-symmetric Markov matrix that appear in
that context has been symmetrized, which can always be accomplished if
detailed balance is satisfied.

Given these basis states, one seeks the ``best'' solution to the
linear variational problem in Eq.~(\ref{eq.varexc}) in the sense that
for all $i$ the \keyword{Rayleigh quotient}
$\bra{\tilde\psi_{i}}|\Ham\ket{\tilde\psi_i}/\bra{\tilde\psi_{i}}\ket{\tilde\psi_i}$
is stationary with respect to variation of the coefficients of the
matrix $D$.  The solution is that the matrix of coefficients
$D_i^{(j)}$ has to satisfy the following generalized eigenvalue
equation
\beq
\sum_{i\,=\,1}^n H_{ki}\,D_i^{(j)} \,=\, {\tilde E}_j
\sum_{i\,=\,1}^n N_{ki}\,D_i^{(j)},
\label{eq.geneig}
\eeq
where
\beq
H_{ki}\,=\,\bra {u_k}|\Ham \ket {u_i},\mbox{ and }N_{ki}\,=\,\bra {u_k}\ket {u_i}.
\eeq

We note a number of important properties of this scheme.  Firstly, the
basis states $\ket{u_{i}}$ in general are not orthonormal.  Secondly,
it is clear that any nonsingular linear combination of the basis
vectors will produce precisely the same results, obtained from the
correspondingly transformed version of Eq.~(\ref{eq.geneig}).  The
final comment is that the variational eigenvalues bound the exact
eigenvalues from above, {\it i.e.,} $\tilde E_i \ge E_i$, where we
assume $E_1\le E_2 \le \dots$.  One recovers {\it exact} eigenvalues
$E_i$ and the corresponding eigenstates, if the $\ket{u_{i}}$ span the
same space as the exact eigenstates, or in other words, have no
admixtures of more than $n$ states.

The required \keyword{matrix element}s can be computed using the
standard variational \MC\ method.
The power method can subsequently be used to reduce the variational
bias.  Formally, one simply defines new basis states
\beq
\ket {u_{i}^{(p)}} \,=\, G^p\,\ket {u_{i}}
\eeq
and substitutes these new basis states for the original ones.  In
quantum mechanical applications, where $G=\exp (-\tau \H)$, the
corresponding matrices
\beq
H_{ki}^{(p)}\,=\,\bra {u_k^{(p)}}|\Ham \ket {u_i^{(p)}}
\eeq
and
\beq
N_{ki}^{(p)}\,=\,\bra {u_k^{(p)}}\ket {{u_i}^{(p)}}
\eeq
can be computed by pure-diffusion \MC .\cite{caffarel} We note that,
\MC\ yields these matrix elements up to an irrelevant overall normalization
constant.

As an explicit example illustrating the nature of the \MC\
time-averages that one has to evaluate in this approach, we write down
the expression for $N_{ij}^{(p)}$ as used for the computation of
eigenvalues of the \keyword{Markov matrix} relevant to the problem of
critical slowing down, discussed in detail in the next section.  One
estimates this matrix as
\beq
N_{ij}^{(p)} \,\propto\, \sum_t {u_i(\S_t) \over \psiB(\S_t)}
{u_j(\S_{t+p}) \over \psiB (\S_{t+p})},
\label{eq.uucorrel}
\eeq
where the $\S_t$ are configurations forming a time series that is
designed to sample the distribution of a system in thermodynamic
equilibrium, {\it i.e.,} the Boltzmann distribution $\psiB^2$.  It
turns out that in this particular case, this distribution, the
dominant eigenstate, has sufficient overlap with the magnitude of the
sub-dominant states so that one can compute all matrix elements
$N_{ij}^{(p)}$ simultaneously without introducing a separate guiding
function\cite{CeperleyBernu88}.

The expression given in Eq.~(\ref{eq.uucorrel}) yields the
$u/\psiB$-auto-correlation function at lag $p$.  The expression for
$H_{ij}^{(p)}$ is similar, and represents a cross-correlation function
involving the configurational eigenvalues of the \keyword{Markov
matrix} in the various basis states.  Compared to the expressions one
usually encounters in applications to quantum mechanical problems,
Eq.~(\ref{eq.uucorrel}) takes a particularly simple form in which
products of fluctuating weights are absent, because one is dealing
with a probability conserving evolution operator from the outset in
this particular problem.

\section{Universal Amplitude Ratios in Critical Dynamics}\index{critical point amplitude}

Before continuing our general discussion, we temporarily change the
topic to introduce stochastic dynamics of critical systems.  What make
such systems interesting, is that one can distinguish universality
classes in which behavior does not depend on many of the microscopic
details.  For static critical phenomena, it is known that universality
classes can be identified by dimensionality, symmetry of the order
parameter, and the range of the interactions.  For dynamical
phenomena, there are additional features such as whether or not the
dynamics is local or subject to conservation laws.

On approach of a critical point, the correlation length $\xi$
diverges. The \keyword{dynamical exponent} $z$ governs the
corresponding divergence of the correlation time $\tau$ by means of
the relation $\tau \,\propto\, \xi^z$.  Since the \keyword{critical
exponent} $z$ is one of the universal quantities, it has been used to
identify \keyword{universality class}es.  Unfortunately, $z$ does not
vary by much from one \keyword{universality class} to another, and
this poses a serious computational problem in terms of the accuracy
required to obtain significant differences.  One of the outcomes of
the work reviewed here is that there are other quantities within
computational reach, namely universal amplitude ratios.\index{critical
point amplitude}\cite{NB98.prl} These ratios may serve as additional,
and possibly more sensitive identifiers of \keyword{universality
class}es.  We shall consider various systems belonging to a single
\keyword{universality class}, and we assume that the representatives
of the class are parameterized by $\kappa$.

If a thermodynamic system is perturbed out of equilibrium, different
thermodynamic quantities relax back at a different rates.  More
generally, there are infinitely many independent \keyword{relaxation
mode}s for a system in the thermodynamic limit.  The \MC\ methods
reviewed here have been used to compute relaxation times of Ising
models on square $L \times L$ lattices at the critical
point.\cite{NB98.prl}

Let us denote by $\tau_{\kappa i}(L)$ the relaxation time of mode $i$
of a system of linear dimension $L$.  As indeed scaling theory
suggests, it turns out that the relaxation time has the following
factorization property
\beq
\tau_{\kappa i}(L)  \,\approx\, m_{\kappa} A_i L^z,
\label{eq.factor.tau}
\eeq
where $m_{\kappa}$ is a {\it non-universal} metric factor, which
differs for different representatives of the same universality class
as indicated; $A_i$ is a {\it universal} amplitude which depends on
the mode $i$; and $z$ is the {\it universal} dynamical exponent
introduced above.

Formulated as a computational problem, one has the following.  Suppose
$\S\,=\,(s_1,...,s_{L^2})$, with $s_i\,=\,\pm 1$, is a spin configuration and
$\rho_{t}(\S)$ is the probability of finding $\S$ at time $t$. The
probability distribution evolves in time according to
\beq
\rho_{t+1}(\S)\,=\,\sum_{\Sp} P(\S|\Sp)\rho_{t}(\Sp).
\eeq
The detailed structure of the \keyword{Markov matrix} $P$ is of no
immediate importance for the current discussion.  All that matters is
that it satisfies \keyword{detailed balance}, has the Boltzmann
distribution $\psiB^2$ as its stationary state.  Also, $P$ is a
single-spin flip matrix, \ie\ $P(\S|\S')$ vanishes if $\S$ and $\Sp$
differ by more than a single spin.  The desired relaxation time of
mode $i$ is given by
\beq
\tau_i(L)\,=\,-L^{-2}/ \ln \lambda_i(L),
\eeq
where $\lambda_i$ is an eigenvalue of \keyword{Markov matrix} $P$.  We
obtained the previous expression by assuming a single-spin flip
Markov matrix, so that the $L^2$ in the denominator produces a
relaxation time measured in units of sweeps, {\it i.e.} flips per
spin.

\section{Trial State Optimization}

To verify Eq.~(\ref{eq.factor.tau}), it is important to obtain
estimates that are exact within the range of the estimated error. For
this purpose we use a set of optimized variational basis functions, to
which we subsequently apply the projection procedure described in
Section 2 to remove the variational bias.

As mentioned, the \MC\ projection increases the statistical noise, and
the solution to this problem is to improve the variational basis
functions.  We shall now discuss how this is done and we consider the
problem using the language of the Schr\"odinger equation.

We first consider the ground state and review how one can optimize a
many-, say 50-parameter trial function $\psiT(R)$.\cite{UWW} The {\it
local energy} $\E(R)$ is defined by
\beq
\H \psiT(R) \,\equiv\, \E(R) \psiT(R).
\eeq
The variance of the local energy is given by
\beq
\chi^2\,=\, \langle (\H-\overline{\E})^2 \rangle\,=\,
{\int |\psiT(R)|^2 [\E(R)-\overline{\E}]^2\,dR\,/\,
\int |\psiT(R)|^2\,dR}.
\label{eq.chisq}
\eeq
A property that we shall exploit later is that $\chi^2\,=\,0$ for any
eigenstate, not just the ground state.

The following sums up the \MC\ optimization procedure for a single
state:
\begin{enumerate}
\item\label{step.sample}
Sample $R_1,\dots,R_s$ from $\psiT^2$ a typical sample size has
$s\,\approx\, 3,000$.
\item
Approximate the integrals in Eq.~(\ref{eq.chisq}) by \MC\ sums.
\item
Minimize $\chi^2$ as follows, while keeping this sample {\it
fixed}. For each member of the sample $R_1,\dots,R_s$:
\item
Compute $\psiT(R_1),\dots,\psiT(R_s)$.
\item
Compute $\H\psiT(R_1),\dots,\H\psiT(R_s)$.
\item
Find $\overline{\E}$ from least-squares fit of
\beq
\H \psiT(R_\sigma) \,=\, \overline{\E} \psiT(R_\sigma),\ \sigma\,=\,1,\dots,s.  \label{eq.HE}
\eeq
\item
Minimize the sum of squared residues of Eq.~\ref{eq.HE}.\footnote{Once
the parameters are changed from the values they had in step
\ref{step.sample}, one should use an appropriately weighted sum of
squared residues.\cite{UWW}}
\end{enumerate}

This procedure can be generalized immediately to a {\it set} of basis
functions, as required to implement Eq.~(\ref{eq.varexc}).  The only
new ingredient is a guiding function $\psig^2$ that has sufficient
overlap with all basis states used in the computation.  For this
purpose one can conveniently use the groundstate raised to some
appropriate power less than unity.

This yields the following algorithm to optimize basis states for $n$
dominant eigenvalues:
\begin{enumerate}
\item\label{step.sample2}
Sample $R_1,\dots,R_s$ from $\psig^2$.
\item
Compute the arrays
\beq
\left(\begin{array}{c}u^{(1)}(R_1)\\u^{(2)}(R_1)\\\vdots\end{array}\right)
,\dots,
\left(\begin{array}{c}u^{(1)}(R_s)\\u^{(2)}(R_s)\\\vdots\end{array}\right).
\eeq
\item
Compute the arrays
\beq
\left(\begin{array}{c}\H u^{(1)}(R_1)\\\H u^{(2)}(R_1)\\\vdots\end{array}\right)
,\dots,
\left(\begin{array}{c}\H u^{(1)}(R_s)\\\H u^{(2)}(R_s)\\\vdots\end{array}\right).
\eeq
\item
Find the \keyword{matrix element}s $\overline{\E}_{ij}$ from the
appropriately weighted least-squares fit to
\beq
\H u^{(i)}(R_\sigma) \,=\, \sum_{j\,=\,1}^n\overline{\E}_{ij} u^{(j)}(R_\sigma),
\ \sigma\,=\,1,\dots,s.
\eeq
\item
Vary the parameters to optimize the fit, as explained below.
\end{enumerate}

In case of a perfect fit, the eigenvalues of the truncated Hamiltonian
matrix $\Em\,=\,(\overline\E_{ij})_{i,j\,=\,1}^n$ are the required
eigenvalues, but in real life one has to optimize the parameters of
the basis functions, which can be done as follows:
\begin{enumerate}
\item
Divide the sample in blocks and compute one Hamiltonian matrix $\Em$
per block.
\item
Minimize the variance of the $\Em$-spectra over the blocks.
\end{enumerate}
The variance vanishes if the basis functions $u^{(i)}$ are {\it linear
combinations} of $n$ eigenstates of $\H$.  This gives rise to a
computational problem, \viz, the variance is near-invariant under
linear transformation of the $u^{(i)}$.  This approximate ``gauge
invariance'' gives rise to near-singular, non-linear optimization
problem.  This can be avoided by simultaneously minimizing the
variance of both the spectrum of the ``local'' Hamiltonian matrix
$\Em$ and the local energy $\E$ of the individual basis functions.

Finally, the variational bias of the eigenvalue estimates obtained
with the optimized basis states is reduced by using \MC\ to make the
substitution discussed previously
\beq
\ket{u^{(i)}} \to e^{-\H \tau}\,\ket{u^{(i)}}.
\eeq
For this purpose, one has to use the short-time approximation of
$\exp(-\H \tau)$.\cite{CeperleyBernu88} To apply the preceding scheme
to the problem of critical dynamics, all one has to do is to make use
of the fact the analog of the quantum mechanical evolution is the
symmetrized Markov $\Phat$ of stochastic dynamics, which is defined as
\beq
\Phat(\S|\Sp) \,=\, {1 \over \psiB(\S)} P(\S|\Sp)  \psiB(\Sp),
\eeq
in terms of which we have the correspondence
\beq
e^{-\H \tau} \to \Phat^t.
\eeq

\section{Xe Trimer: a Test Case}

As an example that illustrates the accuracy one can obtain by means of
the optimization schemes discussed above, we present results for a Xe
trimer interacting via a Lennard-Jones potential. To be precise, we
write the Hamiltonian of this system in {\it reduced} units as
\beq
\H=-{1 \over 2m} \nabla^2+\sum_{i<j}(r_{ij}^{-6}-2) r_{ij}^{-6},
\eeq
where the $r_{ij}$ denote the dimensionless interparticle distances.
We {\it define} Xe to correspond to $m^{-1}=7.8508\times 10^{-5}$,
which probably to four significant figures \cite{constants} agrees
with Leitner {\it et al.}.\cite{Leitner91}

Table \ref{Xe.table} shows results for variational energies of the lowest five
completely symmetric states of a Lennard-Jones Xe trimer.  The results
are compared with results obtained by the discrete variable
representation truncation-diagonalization method.\cite{Leitner91} The
basis functions used in this computation are of the same general form
used in earlier work with an additional polynomial prefactor for
excited states.\cite{MuNi94,MeMuNi96}

Clearly, we obtain consistently lower reduced energies, which we
attribute to lack of convergence of the results of Leitner {\it et
al.}\cite{Leitner99}

\begin{table}[htbp]
\begin{center}
\begin{tabular}{c|l|l|l}
k&  $E_k$      &  $\sigma$   & Leitner {\it et al.}\\
\hline
0& -2.845 241 50 &  1 $\times 10^{-8}$  & -2.844\\
1& -2.724 955 8  &  1 $\times 10^{-7}$  & -2.723\\
2& -2.675 065    &  1 $\times 10^{-6}$  & -2.664\\
3& -2.608 612    &  2 $\times 10^{-6}$  & -2.604\\
4& -2.592 223    &  3 $\times 10^{-6}$  & -2.580\\
\end{tabular}
\end{center}
\caption{Variational reduced energies compared with estimates of Leitner {\it et al.}}
\label{Xe.table}
\end{table}

\section{Critical Point Dynamics: Results}
Next we briefly address the issue of the choice of trial functions for
the eigenstates of symmetrized \keyword{Markov matrix} $\Phat$.  We write
\beq
u(\S)\,=\,f(\S) \times \psiB(\S).
\eeq
For the modes we considered, $f(\S)$ was chosen to be a rotationally
and translationally invariant polynomial in long-wavelength Fourier
components of $\S$, the lowest-order one of which is simply the
magnetization.  Corresponding to the order parameter and energy-like
modes, we considered polynomials either odd or even under the
transformation $\S
\to -\S$.

\begin{figure}[htb]
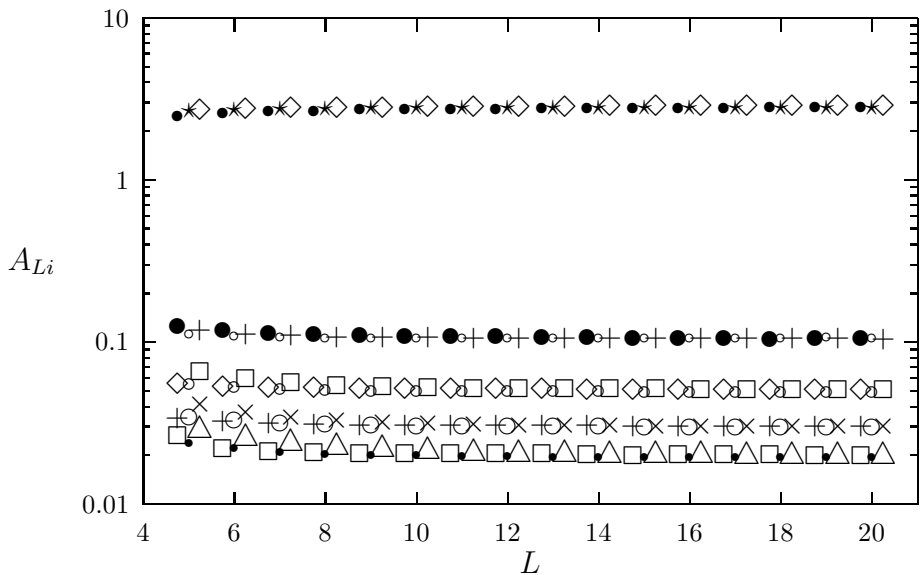

\begin{center}
\input plot
\vspace{4mm}
\caption{
Universality of relaxation-time amplitudes, shown in a plot of the
effective, size-dependent amplitudes $A_{Li}$ on a logarithmic scale.
To separate data points for the three models, the NEQ data were
displaced to the left and the EQN data to the right.  The data
collapse predicted by Eq.~(\ref{eq.factor.tau}) was produced by
fitting the \keyword{metric factor}s of the NN and NEQ models.
Amplitudes of odd and even states alternate in magnitude.}
\label{fig.scaled_gaps}
\end{center}
\end{figure}

We briefly discuss some of the results that illustrate the validity of
Eq.~(\ref{eq.factor.tau}).  Figure~\ref{fig.scaled_gaps}.  shows plots
of the effective amplitudes for the three dominant odd, and two
dominant even modes of three different Ising models on $L\times L$
lattices.  Of the three Ising models we studied, the first one, the NN
model, had nearest-neighbor couplings only. The other two also had
next-nearest-neighbor couplings.  In one of them, the equivalent
neighbor or EQN model, both couplings were of equal ferromagnetic
strengths.  In the third or NEQ model, the nearest-neighbor coupling
was chosen ferromagnetic and of twice the magnitude of the
antiferromagnetic next-nearest-neighbor coupling.

To obtain estimates of the amplitudes of the \keyword{relaxation
mode}s, we fit the computed correlation times to expressions of the
form
\begin{equation}
\tau_{i}(L) \,\approx\, L^z \sum_{k\,=\,0}^{n_{\rm c}} \alpha_{ki} L^{-2k}.
\label{eq.fit}
\end{equation}

In our computation of the non-universal \keyword{metric factor}s, this
quantity was set equal to unity by definition for the EQN model.
Table \ref{tab.metric} shows the \keyword{metric factor}s computed for
each mode separately as the ratio of the computed amplitudes.  In
agreement with the scaling prediction in Eq.~(\ref{eq.factor.tau}),
the computed \keyword{metric factor}s depend only on the model but not
on the mode.
\begin{table}[htb]
\caption{
Non-universal \keyword{metric factor}s $m_{\kappa}$, as defined in
Eq.~(\ref{eq.factor.tau}), computed for the NN and NEQ models. The
modes indicated by o1, o2, and o3 are odd under spin inversion; the
remaining two, e2 and e3, are even.}
\begin{center}
\begin{tabular}{lll}
\hline
    \quad\quad &\quad\quad\ \ NEQ        \quad\quad & \quad\quad\ \ \ NN\\
\hline
o1  \quad\quad & \quad\quad 2.389(1)  \quad\quad & \quad\quad 1.5569  (5)\\
e2  \quad\quad & \quad\quad 2.394(2)  \quad\quad & \quad\quad 1.5569  (5)\\
o2  \quad\quad & \quad\quad 2.393(2)  \quad\quad & \quad\quad 1.5567  (6)\\
e3  \quad\quad & \quad\quad 2.391(2)  \quad\quad & \quad\quad 1.554   (2)\\
o3  \quad\quad & \quad\quad 2.385(4)  \quad\quad & \quad\quad 1.554   (2)\\
\hline
\end{tabular}
\end{center}
\label{tab.metric}
\end{table}

Finally we mention that the spectral gaps of the \keyword{Markov
matrix} vary over a considerable range
\beq
1-\lambda_i(L)\,\approx\, L^{-(d+z)}\,\approx\, L^{-4.17},
\eeq
\ie\, from approximately $3 \times 10^{-3}$ for $L\,=\,4$
to $3\times 10^{-6}$ for $L\,=\,21$. For details of the numerical analysis
based on Eq.~(\ref{eq.fit}) we refer the interested reader to
Ref.\onlinecite{NB98.prl}. Suffice it to mention that the value
obtained for the universal dynamic \keyword{critical exponent} $z$
featured in Eq.~(\ref{eq.factor.tau}) is $z\,=\,2.167\pm0.002$ which is
indistinguishable from 13/6.

\acknowledgments
This work was supported by the (US) National Science Foundation
through Grants DMR-9725080 and CHE-9625498.  It is a pleasure to
acknowledge helpful e-mail exchanges with David Leitner.

\end{document}